# Thin-shell wormholes in Rastall gravity

**Iarley P. Lobo**[1,2,a], **Martín G. Richarte**[3,4,b] , **J. P. Morais Graça**[1,c], **H. Moradpour**[5,d]

[1] Departamento de Física, Universidade Federal da Paraíba, Caixa Postal 5008, João Pessoa, PB 58059-900, Brazil
[2] Departamento de Física, Universidade Federal de Lavras, Caixa Postal 3037, Lavras, MG 37200-000, Brazil
[3] PPGCosmo, CCE - Universidade Federal do Espírito Santo, Vitória, ES 29075-910, Brazil
[4] Departamento de Física, Facultad de Ciencias Exactas y Naturales, Universidad de Buenos Aires, Ciudad Universitaria 1428, Pabellón I, Buenos Aires, Argentina
[5] Research Institute for Astronomy and Astrophysics of Maragha (RIAAM), University of Maragheh, P.O. Box 55136-553, Maragheh, Iran

**Abstract** We constructed thin-shell wormholes using the well-known "cut and paste" technique for static black holes sourced by an anisotropic fluid within the context of the Rastall gravity. Using the generalized Lanczos equations we explored the energy conditions at the wormhole's throat along with the traversability condition. We determined the stability regions when the wormhole configurations are surrounded by different bulk sources and obtained that the stability regions are considerably modified due to the non-trivial dependence of the Rastall parameter in the effective potential energy.

## 1 Introduction

Wormholes are probably one of the most extraordinary theoretical objects ever discovered within the framework of general relativity (GR), as it was firstly noticed by Einstein. Such gravitational configurations are hypothetical passages through different locations of the spacetime that can allow an observer to move between different parts of the Universe, or even between different universes separated by some horizon. In GR, wormholes solutions are supported by exotic matter, i.e., matter that violates at least the weak energy condition (WEC) [1]. Although it is well-known that quantized fields can indeed simulate such kind of exotic matter, this occurs only in some restricted circumstances. Given the fact that one cannot avoid the use of exotic matter to construct wormholes then it would be useful to find a way to at least minimize the amount of exotic matter necessary for their existence. In fact, Visser found a way to restrict the amount of exotic matter to a thin-shell which separates two different regions of the spacetime [2]. These solutions obtained by the aforesaid procedure are known as thin-shell wormholes provided they are constructed from two copies of a black hole solution that are glued together after some portions of its original spacetimes, with all its interior horizon, have been removed. The place where these two copies are glued is called the throat of the wormhole, and contains all the necessary exotic matter necessary to sustain these gravitating wormholes. In his seminal paper, Visser took two copies of the Schwarzschild solution to build a thin-shell wormhole [2,3], but since then, several other spacetimes have been explored in the literature both in GR [4–20], and in other theories of gravity [21–28].

---

[a] e-mail: iarley_lobo@fisica.ufpb.br
[b] e-mail: martin@df.uba.ar (corresponding author)
[c] e-mail: jpmorais@gmail.com
[d] e-mail: h.moradpour@riaam.ac.ir

Rastall gravity is a theory where the covariant derivative of the energy-momentum tensor is not zero [29]. The main idea behind it is that all our laws of conservation based on spacetime symmetries are tested only in the weak-field regime of gravity. Thus, if the covariant derivative of the energy-momentum tensor is proportional to the derivative of the Ricci scalar, then momentum and energy conservation will be recovered as a first approximation in the weak-field regime. In fact, it has been argued that Rastall gravity could be equivalent to Einstein's gravity [30], but we believe that, as long as the energy-momentum tensor used in the theory remains the same as the one defined in Einstein's gravity, the two theories are not equivalent at all (cf. [31] for further discussions on this topic). One of the main features of the Rastall gravity is that its vacuum solutions are the same solutions obtained in GR. Hence, if we want to go further in the analysis of thin-shell wormholes then we must look for black hole solutions surrounded by matter. With the recent discovery that the universe is currently undergoing a phase of accelerated expansion, the idea that the universe is pervaded by some kind of exotic field has been renewed, and therefore it seems appropriate to explore such kind of matter sources as a possible source to sustain thin-shell wormholes.

A black hole solution, surrounded by a particular kind of exotic matter, has been studied by Kiselev [32]. Although the aforesaid research was based on exotic matter, it can be generalized to any kind of matter which can be modeled as an anisotropic fluid. In fact, it has been generalized to several kinds of other fields within the context of Rastall gravity [33]. After that, the thermodynamic properties of those solutions have been studied as well [34]. Besides, some wormhole solutions were obtained within the framework of Rastall gravity recently [35,36].

We are going to construct thin-shell wormholes solutions within the framework of Rastall gravity (RG). To achieve such goal, we will apply the cut-and-past technique to a black hole solution which is surrounded by some fluids. Our aim is to study some properties of these wormholes, such as their energy conditions and their stability. For a such study, we will generalize the junction conditions and Lanczos equation governing the energy-momentum of the thin-shell. In general relativity, the violation of the WEC is compulsory, however, this situation can be avoided within the framework of modified gravity. As we will see, one of the features of Rastall gravity is that wormholes solutions can be constructed with regular matter.

This paper is organized as follows. In Sect. 2, we will review the black hole solution surrounded by an anisotropic fluid within the Rastall gravity. To properly characterize the spacetime geometry we introduce the concept of average equation of state and the idea of partial equation of state. In this way, we will be able to understand whether the matter distribution corresponds to an anisotropic fluid or not. In Sect. 3, we will obtain the junction conditions and Lanczos equations associated with the thin-shell. Later, we will briefly introduce the cut-and-past technique along with some discussion about the traversability condition. We will explore the matter content at the wormhole's throat by studying the violation or not of different energy conditions. In Sect. 4, we will study the regions of stability when the average equation of state mimics dust, radiation, and a cosmological constant even though the bulk fluid is not isotropic in general provided the partial equation of states are different from the average one. In Sect. 5, the conclusions are summarized.

## 2 A black hole surrounded by an anisotropic fluid in RG

Let us begin by obtaining some exact black hole solutions, surrounded by an anisotropic fluid, in Rastall gravity [32,33]. For simplicity, we consider that the manifold can be written as the product $\mathcal{M}_2 \times S^2$, where the line element for $\mathcal{M}_2$ takes the same form of Schwarzschild-like metric, so we will use the following ansatz

$$ds^2 = -f(r)dt^2 + \frac{1}{f(r)}dr^2 + r^2 d\Omega^2, \tag{1}$$

where the solid angle is $d\Omega^2 = d\theta^2 + \sin^2(\theta)d\phi^2$. One reason for choosing the previous ansatz for the metric is that there are several black hole solutions reported within the context of Rastall gravity with this specific ansatz [32,33,37,38]. The time and spatial components of a spherically symmetric energy-momentum tensor that mimics an anisotropic fluid, as studied by Kiselev, are given by

$$T^t_t = T^r_r = -\rho(r) \tag{2}$$

and

$$T^\theta_\theta = T^\phi_\phi = \frac{1}{2}(1+3\omega)\rho(r), \tag{3}$$

where the radial pressure and the tangential pressure are given by $p_r = -\rho$ and $p_\theta = p_\phi = \frac{1}{2}(1+3\omega)\rho(r)$, respectively. Let us make some comments regarding the notation used in this article and the one employed by Visser and collaborators [37,38]. The average pressure is defined as $\bar{p} = [p_r + p_\theta + p_\phi]/3$, the relative anisotropy factor is $\Delta = [p_r - p_\theta - p_\phi]/\bar{p}$ and the equation of state for the average pressure is given by $\omega = \bar{p}/\rho$ [37], therefore we can see that $p_\theta = p_\phi \neq p_r$ and the fluid is not isotropic ($\Delta \neq 0$) neither perfect [37]. Although the radial equation of state satisfies the linear relation $p_r = \omega_r \rho$ with $\omega_r = -1$, the equation of state for the tangential component is completely different, namely $p_\theta = \omega_\theta \rho$ with $\omega_\theta = (1+3\omega)/2$. The latter fact indicates once again that the fluid is not perfect or isotropic. Let us explore how the energy-momentum reads for different values of the average effective equation of state. The average equation of state parameter $\omega$ and the other partial equation states parameter are related to each other in the following way:

$$\omega = \frac{\omega_r + \omega_\theta + \omega_\phi}{3} = \frac{-1 + 2\omega_t}{3}. \tag{4}$$

An average cosmological constant corresponds to $\omega = -1$ ($p_\theta = -\rho = p_r$) so the energy momentum tensor reads $T^\mu_\nu = \rho\,\text{diag}(-1,-1,-1,-1)$. For an average dust-like equation of state ($\omega = 0$) the partial equation of states are $(\omega_r, \omega_\theta, \omega_\phi) = (-1, 1/2, 1/2)$ so the relative anisotropy factor goes to infinity and the energy-momentum tensor can be recast as $T^\mu_\nu = \rho\,\text{diag}(-1,-1,1/2,1/2)$. For an average equation of state with a radiation-like the relative anisotropy factor is non-vanishing ($\Delta = -6$) and the energy-momentum tensor is $T^\mu_\nu = \rho\,\text{diag}(-1,-1,1,1)$ [37].

We must now use the energy-momentum tensor (2)–(3) in Rastall gravity in order to obtain some black hole solutions. As we stated earlier, in the Rastall theory, the covariant derivative of the energy-momentum tensor does not necessarily have to be zero, since conservation of momentum and energy have been tested only in the flat or weak-field gravity regimes [29],

$$\nabla_\mu T^{\mu\nu} = \lambda \nabla^\nu R, \tag{5}$$

where $\lambda$ is the so called Rastall parameter. It is worth mentioning that the Rastall theory leads to GR in the $\lambda \to 0$ limit. We can see that, in Minkowski spacetime, we recover all the well-know laws of conservation of energy and momentum. Recently, the Rastall theory has been the center of several debates. It was claimed that the theory is classically equivalent to general relativity [30]. To show that, Visser argued that the energy-momentum tensor can be re-arranged in order to satisfy the covariant conservation through a process called rastallization. But, Darabi et al. presented several arguments against Visser's claims (cf. [31]). Assuming that the energy-momentum tensors of both gravitational theories should be equivalent leads to the inequivalence of both gravitational theories [31]. In reality, one expects that the non-conservation of the energy-momentum tensor must be accommodated as a non-minimal coupling between matter and gravity (see, for instance [39]). In fact, some authors claimed that the Rastall theory may be in accordance with the Mach principle for a gravity theory in the sense that the local matter configurations depend on the distribution of matter of the whole spacetime [40]. Besides, the statistical analysis based on the galaxy-galaxy strong gravitational lensing indicated that the Rastall's parameter $\beta = \kappa\lambda$ is non-zero but is order of unity [41], supporting the idea that at the galactic scale it is possible to distinguish both theories. Moreover, the Bayesian analysis performed to constrain the rotation curves of 16 low surface brightness spiral galaxies showed that the Rastall's parameter $\beta$ is of order $10^{-1}$ [42], which is consistent with the strong lensing estimation [41]. At the cosmological scale, the supernovae type Ia data associated with the Constitution sample put stronger bounds on the Rastall's parameter, yielding to $\beta \simeq \mathcal{O}(10^{-4})$ [43,44]. However, the latter constrained needs to be updated with the latest cosmological dataset in order to give a clear constraint on $\beta$. Briefly, the range of the Rastall's parameter is not well-constrained in all the scales so one cannot rule out the theory completely and further analysis is required. We should mention that the nonconservation of the energy-momentum tensor appears in unimodular gravity [45], where it was shown that an effective cosmological constant can be derived from another fundamental point of view [45]. Remarkably, it was shown that the Rastall's theory can be obtained within the context of a $\mathcal{F}(R, T)$-modified gravity with non-minimal coupling, where $R$ is the Ricci scalar and $T$ stands for the trace of the Ricci tensor [46]. In fact, a Lagrangian of the form $\mathcal{F}(R, L_m)$, $L_m$ being the matter Lagrangian, can be accommodated as a representation of the Rastall's theory. These facts confirm that Rastall's theory is not equivalent to GR [46].

Let us turn our focus to Rastall gravity field equations. From Eq. (5), we can write the Rastall's equations of gravity as follows

$$H^{\mu}_{\nu} \equiv G^{\mu}_{\nu} + \lambda\kappa\delta^{\mu}_{\nu}R = \kappa T^{\mu}_{\nu}, \quad (6)$$

where $\kappa$ is Rastall's gravitational constant. For simplicity, we will use units in which $\kappa = 1$. To obtain some solutions within this theory, we should solve the set of Eq. (6) for some given energy-momentum tensor (2)–(3). Besides, taking the trace of Eq. (6), we obtain

$$R(4\lambda - 1) = T, \quad (7)$$

which means that a zero trace ($T = 0$) can be obtained in two different ways, we can impose either $R = 0$ or take the special value $\lambda = 1/4$. Nevertheless, the latter option is not a real possibility in the sense that $T$ would be zero by default, so we must stick with the former option, implying that all vacuum solutions in GR are also solutions in RG. On the other hand, the physical scenario where the energy-momentum tensor has non-vanishing trace is quite different provided RG has more constraints than GR, and as result of that some solutions that exist in GR have no counterparts in RG. The main reason is that the set of field equations (6) must be compatible, and sometimes this happens only for $\lambda = 0$.

We will now proceed to solve the field equations (6) for the energy-momentum tensor given by (2) and (3). As mentioned earlier, we are interested in black hole solutions that can be written as a Schwarzschild-like metric, so we will use the ansatz given in Refs. [32,33,37,38]. Plugging (1) into field equations (6), the components $H^t_t = H^r_r = T^r_r = T^t_t$ give the following expression,

$$\frac{1}{r^2}(rf' - 1 + f) - \frac{\lambda}{r^2}(r^2 f'' + 4rf' - 2 + 2f) = \omega\rho, \tag{8}$$

whereas the other $H^\theta_\theta = H^\phi_\phi = T^\phi_\phi = T^\theta_\theta$ components yield to

$$\frac{1}{r^2}\left(rf' + \frac{1}{2}r^2 f''\right) - \frac{\lambda}{r^2}(r^2 f'' + 4rf' - 2 + 2f)$$
$$= \frac{1}{2}(1 + 3\omega)\rho. \tag{9}$$

Here the prime stands for derivative with respect to the radial coordinate. Our next task is to find a way to solve the above equations. We have two field equations (8)–(9) and two undetermined functions $f(r)$ and $\rho(r)$. To deal with this problem, we are going to choose a particular ansatz for the energy density, namely, $\rho(r) = Ar^\beta$, where $A$ and $\beta$ are both constants. In order to ensure the compatibility of the field equations, the following constraints have to be satisfied,

$$\beta = -\frac{3(1+\omega) - 12\lambda(1+\omega)}{1 - 3\lambda(1+\omega)} \tag{10}$$

and

$$A = \frac{3N(1-4\lambda)(\lambda(1+\omega)-\omega)}{(1-3\lambda(1+\omega))^2}. \tag{11}$$

Here $N$ is simply an integration constant. In short, the metric function can be recast as a deformation of the Schwarzschild solution,

$$f(r) = 1 - \frac{2M}{r} - Nr^{-\eta}, \tag{12}$$

where the new parameter is given by

$$\eta = \frac{1 + 3\omega - 6\lambda(1+\omega)}{1 - 3\lambda(1+\omega)}. \tag{13}$$

Equation (12) tells us that the general solution is parametrized by the mass parameter $M$ and the exponent parameter $\eta$, which at the same time involves the equation of state parameter $\omega$ along with the Rastall parameter $\lambda$. In the limit $\lambda \to 0$, we recover the Kiselev solution [32], as expected. Notice that we did not fix the parameter $\omega$ so this extends the possibilities explored by Kiselev [32] by not restricting ourselves to a quintessence field only [33]. The global topology of the manifold will essentially depend on the numbers and the types of horizons that the metric allows (12).

## 3 Thin-shell wormholes

### 3.1 Junction conditions in RG

Before we embark ourselves in the construction of thin-shell wormholes it is mandatory to obtain the so-called junction condition within the framework of RG. To do so, we consider the usual decomposition of the spacetime in a hypersurface $\Sigma$, determined by their orthonormal vector field **n**. The hypersurface can be spacelike, where the unit normal is timelike $(\mathbf{n} \cdot \mathbf{n}) = -1$, or timelike, where the unit normal is spacelike $(\mathbf{n} \cdot \mathbf{n}) = 1$. The null case will not be discussed here. Taking into account the usual definitions of intrinsic and extrinsic curvatures for the hypersurface $\Sigma$, the components of Einstein tensor in Gaussian coordinates can be recast as [47]

$$G^n_n = -\frac{1}{2}\bar{R} + \frac{1}{2}(\mathbf{n} \cdot \mathbf{n})^{-1}(K^i_i K^j_j - K_{ij}K^{ij}), \tag{14}$$

$$G^n_i = (\mathbf{n} \cdot \mathbf{n})^{-1}(K^{\ m}_i - K^j_{j|i}), \tag{15}$$

$$G^i_j = \bar{G}^i_j + (\mathbf{n} \cdot \mathbf{n})^{-1} \left[ -(K^i_j - \delta^i_j K^l_l)_{,n} - K^l_l K^i_j \right.$$
$$\left. + \frac{1}{2}\delta^i_j K^l_l K^m_m + \frac{1}{2}\delta^i_j K_{lm}K^{lm} \right], \tag{16}$$

where $\bar{R}$ and $\bar{G}^i_j$ are the curvature scalar and the Einstein tensor of the hypersurface $\Sigma$, respectively. Besides, the curvature can be decomposed as

$$R = \bar{R} - (\mathbf{n} \cdot \mathbf{n})^{-1}(K^{ij}K_{ij} + K^i_i K^j_j - 2K^i_{i,n}). \tag{17}$$

We then proceed to find junction conditions when spacetime is separated by a hypersurface $\Sigma$. This is similar to the procedure of finding the junction conditions for electric and magnetic fields separated by a surface with charge and current densities. In fact, we define the stress energy-momentum tensor $S^\alpha_\beta$ of $\Sigma$ to be the integral of $T^\alpha_\beta$ with respect to the proper distance, $n$, measured perpendicularly through $\Sigma$

$$S^\alpha_\beta = \lim_{\epsilon \to 0} \left( \int_{-\epsilon}^{+\epsilon} T^\alpha_\beta \, dn \right). \tag{18}$$

For a well-defined hypersurface, its metric, intrinsic and extrinsic curvature should not present any delta function, which implies that after having performed the integration on the tensor $H^\alpha_\beta = G^\alpha_\beta + \lambda\kappa\delta^\alpha_\beta R$ and taking into account the Rastall field equations we arrive at

$$\int_{-\epsilon}^{+\epsilon} H^n_n \, dn = -2\lambda(\mathbf{n} \cdot \mathbf{n})^{-1} \kappa^i_i = S^n_n, \tag{19}$$

$$\int_{-\epsilon}^{+\epsilon} H^n_i \, dn = 0 = S^n_i, \tag{20}$$

$$\int_{-\epsilon}^{+\epsilon} H^i_j \, dn = -(\mathbf{n} \cdot \mathbf{n})^{-1} \left( \kappa^i_j - \delta^i_j \kappa^l_l + 2\lambda\delta^i_j \kappa^l_l \right) = S^i_j, \tag{21}$$

where $\kappa_{ij} \doteq [K_{ij}] := K^+_{ij} - K^-_{ij}$ represents the jump of the extrinsic curvature across the hypersurface, namely the surface energy-momentum generates a discontinuity in the extrinsic curvature. For **n** a spacelike vector, $(\mathbf{n} \cdot \mathbf{n}) = 1$, the Lanczos equations can be written as

$$\kappa^i_j - \delta^i_j \kappa^l_l + 2\lambda\delta^i_j \kappa^l_l = -S^i_j. \tag{22}$$

In addition to (22), we also have a new contribution due to the Rastall parameter and the non-conservation of the energy-momentum tensor

$$2\lambda\kappa^i_i = -S^n_n, \tag{23}$$

which means that there is momentum associated with the surface layer $\Sigma$ flowing out of it, a feature that is absent in GR. Now, let us obtain the new conservation equation for the stress energy-momentum tensor. From the trace of (22) we obtain a constraint between the trace of the extrinsic curvature and the trace of the energy-momentum,

$$2\kappa_l^l = \frac{S_l^l}{1 - 3\lambda}. \tag{24}$$

Replacing the above constraint in (22), we have

$$\kappa_j^i - \delta_j^i \kappa_l^l = -S_j^i - \frac{\lambda}{1 - 3\lambda} \delta_j^i S_l^l. \tag{25}$$

Using the Bianchi identity $G^{\mu\nu}_{;\mu} = 0$, we can derive the following expression [47]

$$(\kappa_j^i - \delta_j^i \kappa_l^l)_{|i} = -[G_j^n], \tag{26}$$

where the symbol '|' stands for the covariant derivative on the hypersurface with metric $g_{ij}$ and $[G_j^n]$ is the jump in the components of the Einstein tensor. Using Rastall field equations (6) we can get

$$G_\nu^\mu = T_\nu^\mu + \frac{\lambda}{1 - 4\lambda} \delta_\nu^\mu T, \tag{27}$$

therefore, fixing $n$ as the contravariant index of the above equation, we have $G_j^n = T_j^n$. In this way, we are able to determine the conservation equation within RG

$$S_{j|i}^i + \frac{\lambda}{1 - 3\lambda} S_{l|j}^l - [T_j^n] = 0, \tag{28}$$

which is a generalization of the well-known expressions derived in [47] and we can notice a modification that it essentially depends on the 3D covariant derivative of the energy-momentum trace, say $S_{l|j}^l$. We are going to consider a thin-shell supported by a fluid surrounded by a continuous matter distribution such that $[T_j^n] = 0$, and therefore the continuity equation can be found by projecting the above expressions in the direction of a normalized observer with velocity $u^i = \delta_0^i$, namely, we have $u_{|j}^i u_i = 0$ and $u_{|i}^i = 2a^{-1}\dot{a}$, where $a$ represents the radius of the wormhole throat(see the following subsection for further details). In fact, these set of equations can be written as

$$S_j^i = \sigma u^i u_j + \mathcal{P}(\delta_j^i + u^i u_j), \tag{29}$$

$$S_{j|i}^i u^j + \frac{\lambda}{1 - 3\lambda} S_{l|j}^l u^j = 0, \tag{30}$$

Taking into account (30), we may write the modified conservation equation as follows

$$\frac{d}{d\tau}(a^2 \sigma) + \mathcal{P} \frac{da^2}{d\tau} + \frac{\lambda}{1 - 3\lambda} a^2 \frac{d}{d\tau}(\sigma - 2\mathcal{P}) = 0. \tag{31}$$

where $\tau$ is the comoving time associated with an observer living on $\Sigma$.

At this point we need to give a physical interpretation of each term in the conservation law (31). To do so, we write it as

$$d\mathcal{U} + d\mathcal{W} + đ\mathcal{Q} = 0, \tag{32}$$

where the internal energy is given by $\mathcal{U} = \mathcal{A}\sigma$ and the area of the wormhole throat is $\mathcal{A} = 4\pi a^2$. The second term represents the work done by the shell's internal force, $d\mathcal{W} = \mathcal{P}d\mathcal{A}$. The third term stands for the flow of heat released and its origin is intrinsically related with the modifications introduced by the Rastall theory [cf. (5), (30)]. The flow of heat can be written as $đ\mathcal{Q} = (\lambda/1 - 3\lambda)\mathcal{A}\mathbf{d}(\sigma - 2\mathcal{P})$. Notice that the sign of $d\mathcal{U}$ is not defined provided $\sigma$ can takes negative or positive values. A similar situation happens for the work term provided the speed of the shell can be positive if it is expanding or negative if it is contracting. The heat flow does not have its sign defined either. If $0 < \lambda < 1/3$ and $\mathbf{d}(\sigma - 2\mathcal{P}) > 0$ the heat is released but the situation is reversed in the several cases, 1.—$\lambda < 0$ and $\mathbf{d}(\sigma - 2\mathcal{P}) > 0$, 2.—$\lambda > 1/3$ and $\mathbf{d}(\sigma - 2\mathcal{P}) > 0$, 3.—$0 < \lambda < 1/3$ and $\mathbf{d}(\sigma - 2\mathcal{P}) < 0$. Anyway, there is something interesting about the conservation law (32) in the sense that in systems with spherical symmetry (1) $đ\mathcal{Q}$ is generally zero and becomes non-zero if there is a exchange of momentum/energy from the shell to the bulk or vice versa [17–20]. Notice that a wormhole throat with zero pressure and energy density is possible due to the non-linearity of the generalized junction conditions (22) and the conservation equation (32) is satisfied as well [28].[1]

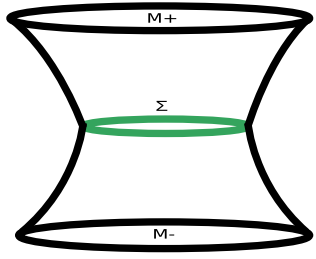

**Fig. 1** Schematic representation for the wormhole geometry obtained after performing the cut and paste procedure to an initial manifold that could have singular regions. The wormhole throat is located at the shell called $\Sigma$ defined by the relation $r - a(\tau) = 0$. The new manifold resulting from gluing $M_+$ with $M_-$ is geodesically complete where the boundaries were identified as $\partial M_- = \partial M_+ = \Sigma$. This representation emphasizes the fact that across the shell the derivatives of the metric exhibit some discontinuities

3.2 Cut and paste method

One way to construct thin-shell wormholes is by using the so-called cut-and-past technique of two spacetimes, usually black hole solutions, but not exclusively, are glued together at a hypersurface, say $\Sigma$, where the throat of the wormhole is located. The latter procedure guarantees that the amount of exotic matter is strictly located at $\Sigma$ [1]. Historically, Visser applied the previous procedure to a Schwarschild solution provided it was the simplest case that one can consider [1–3]. It was also indicated that the stability analysis can be carried out without major issues adopting the cut-and-past technique or thin-shell formalism [2, 3], whereas the analysis of the dynamical stability may not be so straightforward in other approaches [48–50]. In this section we will briefly present the cut-and-past technique. Let us consider the Schwarzschild-like metric given by (1), and take two copies of the same spacetime (see Fig. 1). In doing so, we define the following sub-manifolds

$$M^{(\pm)} \equiv \left\{ r^{(\pm)} \geq a, |a > r_h \right\}, \tag{33}$$

in which $a$ is chosen to be greater than the outer horizon $r_h$, namely, the largest positive value of $r = r_h$ which satisfies the equation $f(r) = 0$. Each copy of the original manifold is geodesically incomplete, with a boundary given by

$$\partial M^\pm \equiv \left\{ r^\pm = a | a > r_h \right\}, \tag{34}$$

---

[1] In fact we will investigate the behavior of these solitonic shells somewhere else in the near future.

but the identification of these boundaries results in a geodesically complete manifold, constructed from the union of the two copies, where the throat of the wormhole is identified with the region $\partial M^+ = \partial M^-$. In accordance with the Darmois–Israel formalism, we can now determine the energy-momentum of the throat [51]. First, let us define our coordinate systems. The spacetime generated by the wormhole is covered by the coordinates $x^\alpha = (t, r, \theta, \phi)$, and the thin-shell throat is defined by the hypersurface

$$\Sigma: R(r, \tau) = r - a(\tau) = 0, \tag{35}$$

where $\tau$ is the proper time along the throat. For static wormholes $a$ should be a constant. However, one way to study the stability of wormholes is to perform a small perturbations on it, so it is convenient to consider that the throat can vary with time. The induced metric on the thin-shell throat can be deduced by taking into account the restriction of the original metric to an hypersurface,

$$ds_\Sigma^2 = -d\tau^2 + a(\tau)^2(d\theta^2 + \sin^2(\theta)d\phi^2). \tag{36}$$

In order to get the energy-momentum tensor on the throat, we must integrate the field equations in a box covering such region, and take the limit where this collapses to a thin-shell. A detailed explanation of this procedure can be found on [47]. We derived those expressions (Lanczos equations) in the previous subsection for RG (cf. 22). The extrinsic curvature is given by

$$K_{ij}^{(\pm)} = -n_\mu^{(\pm)} \left( \frac{\partial^2 x^\mu}{\partial \xi^i \partial \xi^j} + \Gamma_{\alpha\beta}^\mu \frac{\partial x^\alpha}{\partial \xi^i} \frac{\partial x^\beta}{\partial \xi^j} \right)_\Sigma, \tag{37}$$

where $n_\mu$ is the unit vector normal to the throat, $\xi^i$ represents a local coordinate system over the hypersurface different from the one used in the bulk spacetime $x^\beta$, and the superscript indices $(\pm)$ indicate the direction of the normal vector.[2] The Christoffel symbols should also be calculated in relation with each original spacetime, but since we started with two identical copies, we do not have to worry about this point. The unit vector $n_\mu^{(\pm)}$ is given by

$$n_\mu^{(\pm)} = \pm \left( \left| g^{\alpha\beta} \frac{\partial R}{\partial x^\alpha} \frac{\partial R}{\partial x^\beta} \right|^{-1/2} \frac{\partial R}{\partial x^\mu} \right)_\Sigma. \tag{38}$$

For the metric (1), the components of extrinsic curvature are

$$K_\tau^\tau = \pm \frac{\frac{f'}{2} + \ddot{a}}{\sqrt{f + \dot{a}^2}} \quad \text{and} \quad K_\theta^\theta = K_\theta^\theta = \pm \frac{1}{a}\sqrt{f + \dot{a}^2}, \tag{39}$$

where the prime indicates derivative with respect to the radial coordinate, the dot stands for derivative with respect to the proper time, and the functions are all evaluated at the throat ($r = a(\tau)$). In addition to that, the thin-shell intrinsic energy-momentum tensor can be calculated by means of the Lanczos equation (22) using that $S_j^i = (-\sigma, \mathcal{P}, \mathcal{P})$. The surface energy density is

$$\begin{aligned}\sigma &= -(1 - 2\lambda)(\kappa_\theta^\theta + \kappa_\phi^\phi) + 2\lambda \kappa_\tau^\tau \\ &= -\frac{4}{a}\sqrt{f + \dot{a}^2}(1 - 2\lambda) + 2\lambda \frac{f' + 2\ddot{a}}{\sqrt{f + \dot{a}^2}}\end{aligned} \tag{40}$$

---

[2] Note that in our definition of the extrinsic curvature is minus the one used in [47].

whereas the tangential pressure is

$$\mathcal{P} = \left[(\kappa_\theta^\theta + \kappa_\tau^\tau) - 2\lambda(\kappa_\tau^\tau + \kappa_\theta^\theta + \kappa_\phi^\phi)\right]$$
$$= \frac{(2\ddot{a}a + f'a)(1 - 2\lambda) + 2(f + \dot{a}^2)(1 - 4\lambda)}{a\sqrt{f + \dot{a}^2}}. \tag{41}$$

To solve the above field equations, one must choose an equation of state $\mathcal{P} = \mathcal{P}(a)$. However, instead of solving such equations, we will study some properties of these wormholes such as their energy conditions and stability regions.

3.3 Traversability and flare-out condition

To illustrate the main traits of wormhole geometries we look at the flare-out condition [50, 52–54]. Wormhole throat must be a minimal surface in order to guarantee that a physical observer can pass through otherwise the surface may pinch off and disappear. There are several definitions depending on their topological traits [50,52–54]. Let us consider a local system of coordinates over $\Sigma \simeq \mathfrak{R} \times \Sigma_\tau$ given by $(\tau, z^d)$ where $d = (\theta, \phi)$. From (36) we obtain the induced metric on $\Sigma$: $h_{ij} = \text{diag}(-1, \mathcal{H}, \mathcal{H}(a)\sin^2\theta)$ being $\mathcal{H} = a^2$. The induced metric on the spatial section $\Sigma_\tau$ is $\gamma_{dc} = \text{diag}(\mathcal{H}(a), \mathcal{H}(a)\sin^2\theta)$ and its determinant is denoted as $\gamma = \text{det}\gamma_{dc}$. The standard area definition for the wormhole throat is

$$\mathcal{A}(\Sigma_\tau) = \int_{\Sigma_\tau} d^2z\sqrt{|\gamma|}. \tag{42}$$

A wormhole throat acts as a tunnel, in the sense that is traversable, if $\mathcal{A}(\Sigma_\tau)$ satisfies the following conditions [48,49]: (1) Extremality $\delta\mathcal{A}(\Sigma_\tau) = 0$ and (2) Convexity $\delta^2\mathcal{A}(\Sigma_\tau) \geq 0$. It turned out that $\delta\mathcal{A}(\Sigma_\tau)$ is proportional to $\text{Tr}(\kappa_c^d)$ calculated in terms of $\gamma_{dc}$. However, $\delta^2\mathcal{A}(\Sigma_\tau)$ involves the sum $\partial_n\text{Tr}(\kappa_c^d) - [\text{Tr}(\kappa_c^d)]^2$ [52]. So, if the extremality condition is satisfied, $\text{Tr}(\kappa_c^d) = 0$, the minimality condition requires that $\partial_n(\text{Tr}(\kappa_c^d)) \leq 0$, where $\partial_n$ is the derivative along the normal. However, thin-shell wormholes are supported by some matter content then $\text{Tr}(\kappa_c^d) = -\sigma$ does not vanish [54]. To generalize the notion of a wormhole's throat we should retain the minimal area condition $[\partial_n(\text{Tr}\kappa_c^d)) \leq 0]$ but avoid the extremality condition, which can be replaced by $\text{Tr}(\kappa_c^d) < 0$ or $\text{Tr}(\kappa_c^d) > 0$ [54].

Let us focus on static thin-shell wormhole, namely $\dot{a} = \ddot{a} = 0$, within the context of RG. The trace of the extrinsic curvature yields

$$\text{Tr}(\kappa_c^d) = \sqrt{f(a)}\frac{\partial}{\partial a}\left(\ln \mathcal{H}^2\right). \tag{43}$$

Hence, the sign of (43) depends on the sign of $\mathcal{H}'(a)$ for $\mathcal{H} \geq 0$. In RG, we must have $\text{Tr}(\kappa_c^d) > 0$ provided $\mathcal{H}'(a) > 0$ as it happens in GR. To see that, we link $\text{Tr}(\kappa_c^d)$ with $\sigma$ by using the generalized junction conditions (22). Using that $\kappa_j^j = \kappa_\tau^\tau + \text{Tr}(\kappa_c^d)$ along with (41), we arrived at the relation $2(1 - 2\lambda)\kappa_\tau^\tau = 2\mathcal{P} - (1 - 4\lambda)\text{Tr}(\kappa_c^d)$. Hence, $\text{Tr}(\kappa_c^d)$ can be written in terms of the energy density, the pressure, and the Rastall's parameter:

$$\text{Tr}(\kappa_c^d) = \frac{2\lambda\mathcal{P} + \sigma(2\lambda - 1)}{1 + \lambda}. \tag{44}$$

GR is a fixed point $\lambda = 0$ of (44) and it leads $\text{Tr}(\kappa_c^d) = -\sigma$. However, both theories do not agree even at the classical level for any other value of $\lambda$. In fact, the sign of $\text{Tr}(\kappa_c^d)$ depends on the relation between $\sigma$ and $\mathcal{P}$. If we choose $\mathcal{P} = \gamma_b\sigma$ with $\gamma_b$ with $0 \leq \gamma_b \leq 1$, the 2D

trace reads

$$\text{Tr}(\kappa_c^d) = \sigma \left( \frac{-1 + 2\lambda\gamma_b}{1+\lambda} \right). \tag{45}$$

If $\sigma > 0$ then $\text{Tr}(\kappa_c^d) > 0$ for $\lambda > -1$ and $2\lambda\gamma_b > 1$ but is negative defined for $\lambda > -1$ and $2\lambda\gamma_b < 1$ or $\lambda < -1$ and $2\lambda\gamma_b > 1$. If $\gamma_b^{\text{fixed}} = 1/2\lambda > 0$ then the extremality condition is achieved in RG, which is forbidden in GR for thin-shell wormholes. If we first take $\gamma_b^{\text{fixed}} = 0$, and then, we take the limit $\lambda \to \infty$, the extremality condition is reached without imposing $\sigma = 0$. To link the previous results with the standard flare-out condition, we write down $\mathcal{A}(\Sigma_\tau) = 4\pi\mathcal{H}(a)$. A minimal surface is obtained if $\mathcal{A}'(\Sigma_\tau) > 0$ which implies $\mathcal{H}'(a) > 0$. Another useful measure of the openness of the wormhole's throat involves the perimeter given by $\mathcal{P}_\tau = 2\pi\sqrt{\mathcal{H}(a)}$ [53], where $\partial_a \ln \mathcal{P}_\tau = \partial_a [\ln \mathcal{H}^{1/2}] > 0$. Hence, thin-shell wormholes in RG satisfy both flare-out conditions.

3.4 Energy conditions

As is well known the matter content of a field theory can be classified according to several energy conditions. If matter satisfies all these conditions, it is denoted ordinary matter, otherwise is called exotic matter. Here we are not going to study the energy-conditions related with the bulk fields because they have been studied in [33], and our aim is to focus on the energy conditions associated with the matter located on the throat. Due to the spherical symmetry of the problem, the energy-momentum of the matter on the throat has only two independent components to determine, that is, the surface energy density $\sigma$ and the surface pressures $\mathcal{P}_\theta = \mathcal{P}_\phi = \mathcal{P}$. In our coordinate system, the *weak energy condition* (WEC) is satisfied if $\sigma \geq 0$ and $\sigma + \mathcal{P} > 0$ whereas it is violated otherwise. As we can see from Eq. (40), this condition depends on the value of the parameter $\lambda$; it is clearly violated for $\lambda = 0$ (GR). The *null energy condition* (NEC) is satisfied if $\sigma + \mathcal{P} > 0$, and the *strong energy condition* (SEC) holds if $\sigma + \mathcal{P} > 0$ and $\sigma + 2\mathcal{P} > 0$ are both satisfied. Clearly, if matter satisfies the SEC, it also fulfills the NEC. The violation of the NEC and SEC are not obvious, since it depends on the relation between the surface energy density and the surface pressure. For this reason, we must study each case separately, and the violation of the energy-conditions will depend on the radius of the throat. Nevertheless, we must keep in mind that none of these energy conditions are sacred in the sense that they are mostly satisfied by simple source matter but there are reasonable physical models that violate any one of these energy conditions, either classically or at the quantum level, as an example we can think of the negative Casimir vacuum energy densities. In short, WEC cannot be satisfied for spherically symmetric thin-shell wormholes in GR provided it requires exotic matter to support them. However, wormholes with cylindrical symmetries within the context of GR sourced by an anisotropic fluid do not require the violation of WEC [55,56].

Let us consider static thin-shell wormholes in RG and evaluate the constraints coming from the energy conditions with more detail. For instance, the positivity of the energy density implies

$$\lambda a f'(a) - 2f(1-2\lambda) \geq 0. \tag{46}$$

The inequality $\sigma + \mathcal{P} > 0$ leads to simple relation which does not involve the $\lambda$ parameter explicitly

$$a f'(a) \geq 2f, \tag{47}$$

whereas $\sigma + 2\mathcal{P} > 0$ implies

$$af'(a)(1 - \lambda) - 4f \geq 0. \tag{48}$$

Before embark us in a numerical analysis, we are going to explore the previous constraints (46)–(48) from another point of view. To do so, we assume that $f(a) > 0$ which is quite reasonable provided the wormhole manifold should not have singular points such horizons. We separate our analysis in two cases depending on the sign of $\lambda$. The first case corresponds then to $f(a) > 0$ and $\lambda > 0$. Equation (47) implies the NEC is satisfied as long as $af'/2f > 0$ provided $f > 0$. The WEC can be guaranteed if the later condition holds along with the requisite (46): $(af'/2f) \geq (-2+\lambda^{-1})$. But we know that $af'/2f > 0$ so the former condition implies that $\lambda \in (0, 1/2)$. On the other hand, the SEC holds if (47) and (48) are both met. Equation (48) tells us that $(af'/2f)(1-\lambda) \geq 2\lambda$. The constraint $(af'/2f) \geq 2\lambda/(1-\lambda) \geq 0$ dictates that SEC holds if $\lambda \in (0, 1)$. On the other hand, SEC is held if the following conditions remain valid $af'/2f > 0$ and $(af'/2f)(1-\lambda) \geq 2\lambda$ with $1 - \lambda < 0$, which in turn implies that $2\lambda/(1-\lambda) \geq 0$. However, the later result leads to a contradiction because it is not possible to have $\lambda > 1$ and $\lambda < 0$ at the same time. We study the other case with $f > 0$ and $\lambda < 0$. Once again, we have that NEC is satisfied provided $\sigma + \mathcal{P} > 0$ [$af'/2f > 0$] does not involve $\lambda$ explicitly, expect for the index $\eta$ which appears in $f(a)$. WEC cannot be satisfied as long as the conditions $af'/2f > 0$ and $\lambda < 0$ are both forced. Indeed, the positivity of $\sigma$ gives $(af'/2f)(-|\lambda|) \geq (1+2|\lambda|) > 0$ which leads to a contradiction. However, the story is completely different in the case of SEC. We know that the following condition $af'/2f > 0$ is true so the second requisite (48) can be written as $(af'/2f)(1+|\lambda|) + 2|\lambda| > 0$, which is clearly satisfied because each single term is positive defined.

We are going to explore what happens with the ECs when only the Rastall contributions in the generalized junction condition are considered [cf. (22)]. For static thin-shell wormholes, the trace of the extrinsic curvature reads

$$\text{Tr}(\kappa_j^i) = \frac{4\sqrt{f}}{a}\left(1 + a\frac{f'}{4f}\right). \tag{49}$$

The Rastall energy-momentum tensor on the wormhole throat is $\mathbb{S}_j^i = -2\lambda \text{Tr}(\kappa_j^i)$ and we obtain that the energy density and tangential pressure are given by $\sigma_{\text{RG}} = 2\lambda \text{Tr}(\kappa_j^i)$ and $\mathbb{P}_{\text{RG}} = -2\lambda \text{Tr}(\kappa_j^i)$, respectively. For $\lambda > 0$ we arrived at $\sigma_{\text{RG}} > 0$ as long as the condition $af'/f > 0$ remains valid (49). Further, the pressure acquires a simple relation in terms of the energy density, $\mathbb{P}_{\text{RG}} = -\sigma_{\text{RG}}$. As a result of the latter fact we obtain that the equation of state for the wormhole's throat is similar to a running cosmological constant provided $\sigma_{\text{RG}}(a)$ is not constant. More importantly, the null and weak energy conditions are trivially satisfied. As we expected, SEC is violated provided $\mathbb{P}_{\text{RG}} + \sigma_{\text{RG}} = 0$ and $2\mathbb{P}_{\text{RG}} + \sigma_{\text{RG}} = -\sigma_{\text{RG}}$.

The scenario in which the Rastall's parameter is negative definite, we have that NEC is not violated because the relation $\sigma_{\text{RG}} = -|\sigma_{\text{RG}}|$ holds and the pressure behaves as a tension, $\mathbb{P}_{\text{RG}} = |\sigma_{\text{RG}}| > 0$. Although the WEC is violated the SEC is not, namely $\mathbb{P}_{\text{RG}} + \sigma_{\text{RG}} = 0$ and $2\mathbb{P}_{\text{RG}} + \sigma_{\text{RG}} = |\sigma_{\text{RG}}|$. The moral of the previous analysis is that the Rastall contributions give some results that are not possible within the pure GR, and therefore it is clear that at the classical level both theories are not equivalent for a general kind of matter content.

We carry on with our main analysis on the ECs. It is useful to write NEC in terms of dimensionless quantities as follows

$$\overline{\sigma + \mathcal{P}} = \frac{\left(-1 + \frac{3}{2x} + n(1+\frac{\eta}{2})x^{-\eta}\right)}{x\sqrt{1 - \frac{1}{x} - nx^{-\eta}}} \geq 0, \tag{50}$$

where $\overline{\sigma + \mathcal{P}} = M(\sigma + \mathcal{P})$ and the integration constant is set as $N = n(2M)^\eta$ without lost of generality being $n$ a dimensionless constant. Here the dimensionless variable is given by $x = a/2M$. In the case of SEC, the requisite (50) is not enough and it must be supplemented by an extra condition, so it reads

$$\overline{\sigma + 2\mathcal{P}} = \frac{(1-\lambda)\left[\frac{1}{x} + n\eta x^{-\eta}\right] - 4\lambda\left[1 - \frac{1}{x} - nx^{-\eta}\right]}{x\sqrt{1 - \frac{1}{x} - nx^{-\eta}}} \geq 0, \tag{51}$$

where $\overline{\sigma + 2\mathcal{P}} = M(\sigma + 2\mathcal{P})$. Besides the condition (50), we must demand that the positivity of energy density to meet the WEC. The latter condition yields

$$\overline{\sigma} = \frac{-2(1-2\lambda)\left[1 - \frac{1}{x} - nx^{-\eta}\right] - (\lambda/2)\left[\frac{1}{x} + n\eta x^{-\eta}\right]}{x\sqrt{1 - \frac{1}{x} - nx^{-\eta}}} \geq 0, \tag{52}$$

in which $\overline{\sigma} = M\sigma$. Notice that we have the condition of $f(x) > 0$ because it appears under the square root sign.

We must verify that the usual expressions for GR are recovered by taking the limit $\lambda \to 0$. We can double-check that above expressions are correct by considering the limits $\lambda \to 0$ and $n \to 0$. The dimensionless quantities $\overline{\sigma}$, $\overline{\sigma + \mathcal{P}}$, and $\overline{\sigma + 2\mathcal{P}}$ are given by

$$\overline{\sigma} = -\frac{2}{x}\sqrt{1 - \frac{1}{x}} < 0, \tag{53}$$

$$\overline{\sigma + \mathcal{P}} = \frac{-1 + \frac{3}{2x}}{x\sqrt{1 - \frac{1}{x}}}, \tag{54}$$

$$\overline{\sigma + 2\mathcal{P}} = \frac{\frac{1}{x}}{x\sqrt{1 - \frac{1}{x}}}, \tag{55}$$

Equations (53)–(55) agree with the ones reported by Poisson and Visser within the context of GR [57].

3.5 Numerical analysis for the energy conditions

For illustrative purposes, we are going to examine our general previous statements with the help of some numerical analyses. In general, we will look for thin-shell wormholes with radius $x_0 > x_h$ that satisfy the aforesaid energy conditions. Such an analysis will essentially depend on several parameters, namely $n$, $\lambda$, $\omega$. To be more specific, we will explore ECs for different values of the average equation of state parameter, say dust, radiation (extremal and non-extremal) and cosmological constant. Above all, we will show that there are some ranges of the Rastall parameter such that the ECs can be preserved and by doing so we will be able to show the interval in which the wormhole radius lives.

3.5.1 Average dust case $\omega = 0$: anisotropic fluid with $\omega_r = -1$ and $\omega_t = 1/2$

We consider the case of a wormhole surrounded by an average dust fluid, i.e., with equation of state parameter $\omega = 0$ which is anisotropic provided the partial equations of state are $\omega_r = -1$ and $\omega_t = 1/2$. It is useful to redefine the radial coordinate as $x = a/2M$ and by doing so the integration constant of the original metric is fixed as $N = (2M)^\eta$, namely, the

**Table 1** It is displayed the intervals in which NEC, SEC, and WEC are satisfied for the average dust-like fluid

| $\lambda$ | NEC | SEC | WEC |
|---|---|---|---|
| 0 | $2 \leq x_0 \leq 3$ | $2 \leq x_0 \leq 3$ | $\nexists x_0$ |
| 0.15 | $4.29 \lesssim x_0 \lesssim 6.62$ | $4.29 \lesssim x_0 \lesssim 6.62$ | $4.29 \lesssim x_0 \lesssim 4.76$ |
| $-0.15$ | $1.83 \lesssim x_0 \lesssim 2.71$ | $1.83 \lesssim x_0 \lesssim 2.71$ | $\nexists x_0$ |

We set the dimensional wormholes radius as $x_0 = a_0/2M$ whereas the integration constant is given by $N = (2M)^\eta$

metric coefficient reads $f(x) = 1 - x^{-1} - x^{-\eta}$. Table 1 displays the range of validity of the energy conditions for different values of $\lambda$. As a fiducial positive or negative value for the Rastall parameter we select $\lambda = \pm 0.15$.

We obtain that all the energy condition can be preserved for a Rastall parameter $\lambda = 0.15$ (cf. Table 1), but the situation is less favorable for a negative Rastall parameter with $\lambda = -0.15$ in the sense that only WEC and SEC are satisfied. We confirm that RG allows thin-shell wormholes supported by normal matter (cf. Table 1). To be more precise, a negative Rastall parameter reduces the range of possible wormhole radius for which the NEC and SEC are valid in comparison to GR, and also it does not contribute to a validation of the WEC. On the other hand, a positive Rastall parameter allows a wider range of the wormhole radius, in fact, it also helps to preserve WEC. A positive $\lambda$ leads to the existence of the wormhole without the need of violations of all these energy conditions, contrary to what happens in GR. We depicted the energy conditions in terms of the wormhole radius $x_0$ (see Fig. 2).

### 3.5.2 Extremal and non-extremal case for an average radiation-like fluid with $\omega = 1/3$: anisotropic fluid with $\omega_r = -1$ and $\omega_t = 1$

We now move on to another case with an average radiation-like fluid ($\omega = 1/3$) associated with partial equations of state $\omega_r = -1$ and $\omega_t = 1$. In the extremal case the original manifold has only one horizon but the non-extremal one possesses two horizons. We redefine the radial coordinate as $x = a/2M$ and the metric coefficient reads $f(x) = 1 - x^{-1} + \frac{1}{4}x^{-\eta}$ for $N = -(2M)^\eta/4$, the horizon is at $x_h = 0.5$ for any value of the Rastall parameter. In the non-extremal one, the metric coefficient reads $f(x) = 1 - x^{-1} + \frac{1}{5}x^{-\eta}$ for $N = -(2M)^\eta/5$, where the *outer* horizon is at $x_h \approx 0.72$ for any value of $\lambda$. In both cases, the value of $\sigma + \mathcal{P}$ is not affected by the Rastall parameter, we then draw only one curve in top panel of Figs. 3 and 4 (Tables 2, 3).

### 3.5.3 Average cosmological constant-like fluid case with $\omega = -1$: isotropic fluid with $\omega_r = -1$ and $\omega_t = -1$

The last example is an average equation of state parameter $\omega = -1$. The original black hole is surrounded by a cosmological constant fluid with a isotropic distribution of matter, leading a metric of the Schwarzschild–de Sitter kind. We define $x = a/2M$ and choose $N = (2M)^\eta/10$ then the metric coefficient becomes $f(x) = 1 - x^{-1} - \frac{1}{10}x^{-\eta}$. The black hole horizon is located at $x_h \approx 1.15$ whereas the cosmological horizon is placed at $x_h \approx 2.42$. Hence, the wormhole radius must be small provided both horizons are considerably closer to each other.

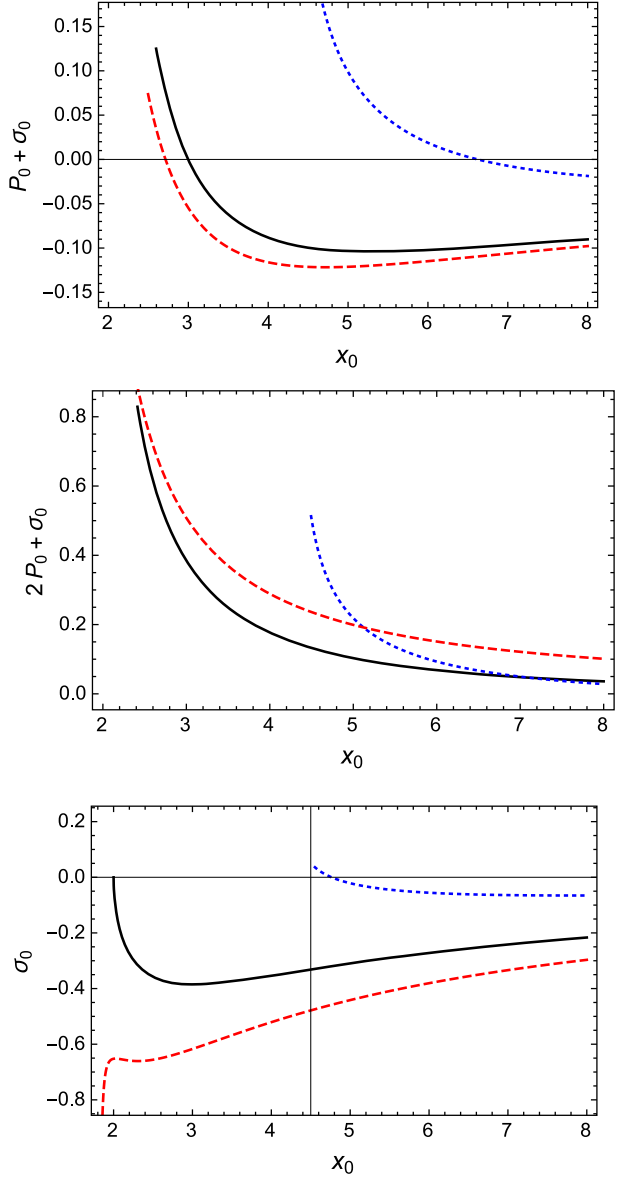

**Fig. 2** It is displayed $\sigma_0 + \mathcal{P}_0$, $\sigma_0 + 2\mathcal{P}_0$ and $\sigma_0$ (in units of $M$) against the dimensionless wormhole radius, $x_0 \doteq a_0/2M$, for the dust fluid case, respectively. It is fixed $N = (2M)^\eta$. The black solid line corresponds $\lambda = 0$, blue dotted line stands for $\lambda = 0.15$, and red dashed line indicates the case with $\lambda = -0.15$. The horizons of the original metric for the average dust fluid are located at $x_h = 2$ for $\lambda = 0$, $x_h \simeq 4.29$ for $\lambda = 0.15$, and $x_h \simeq 1.83$ for $\lambda = -0.15$

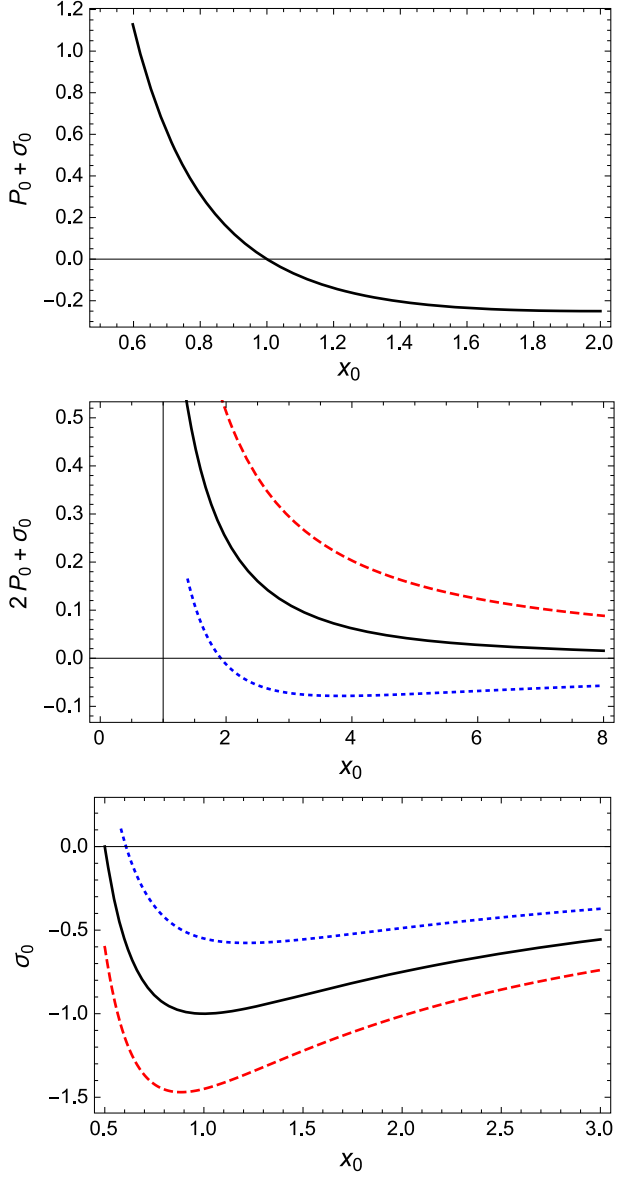

**Fig. 3** Plots of $\sigma_0 + \mathcal{P}_0$, $\sigma_0 + 2\mathcal{P}_0$ and $\sigma_0$ (in units of $M$) against $x_0 \doteq a_0/2M$ for the *extremal average radiation-like fluid*, respectively. We set $N = -(2M)^\eta/4$. There are three different cases: 1. $\lambda = 0$ (black, solid line), 2. $\lambda = 0.15$ (blue, dotted line), and 3. $\lambda = -0.15$ (red, dashed line)

We explored the wormhole radius range in which the energy conditions remain valid. The analysis would seem to indicate that the NEC and SEC are unaffected by the Rastall parameter (regardless the sign adopted), but the WEC remains valid for a positive value of $\lambda$ as is described in Table 4. We draw only one curve in the first case provided NEC is not affected by $\lambda$, see Fig. 5. Although $\sigma_0 + 2\mathcal{P}_0$ ends up being modified by $\lambda$, the SEC is unaffected by the Rastall parameter. The region of validity of the WEC can be explicitly seen in Fig. 5. In brief, RG seems to improve the situation in comparison to GR for positive values of the Rastall parameter in the sense that ECs are not violated.

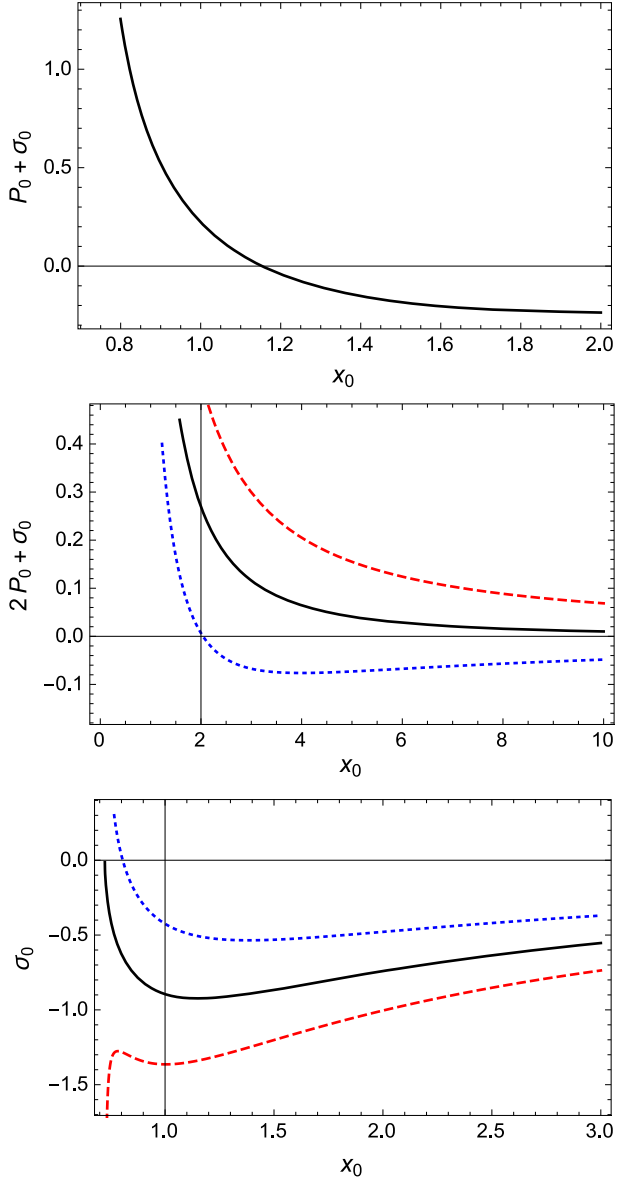

**Fig. 4** Plots of $\sigma_0 + \mathcal{P}_0$, $\sigma_0 + 2\mathcal{P}_0$ and $\sigma_0$ (in units of $M$) against $x_0 \doteq a_0/2M$ for the *non-extremal average radiation-like fluid*, respectively. We set $N = -(2M)^\eta/5$. It is mentioned three cases: 1. $\lambda = 0$ (black, solid line), 2. $\lambda = 0.15$ (blue, dotted line), and 3. $\lambda = -0.15$ (red, dashed line)

**Table 2** Range of validity of NEC, SEC and WEC for the *extremal average radiation-like fluid*

| $\lambda$ | NEC | SEC | WEC |
|---|---|---|---|
| 0 | $0.5 \leq x_0 \leq 1$ | $0.5 \leq x_0 \leq 1$ | $\nexists x_0$ |
| 0.15 | $0.5 \leq x_0 \leq 1$ | $0.5 \leq x_0 \leq 1$ | $0.5 \leq x_0 \lessapprox 0.60$ |
| $-0.15$ | $0.5 \leq x_0 \leq 1$ | $0.5 \leq x_0 \leq 1$ | $\nexists x_0$ |

We set $x_0 = a_0/2M$ and $N = -(2M)^\eta/4$

**Table 3** Range of validity of NEC, SEC and WEC for the *non-extremal average radiation-like fluid*

| $\lambda$ | NEC | SEC | WEC |
|---|---|---|---|
| 0 | $0.72 \lesssim x_0 \leq 1.15$ | $0.72 \leq x_0 \lesssim 1.15$ | $\nexists x_0$ |
| 0.15 | $0.72 \lesssim x_0 \leq 1.15$ | $0.72 \lesssim x_0 \lesssim 1.15$ | $0.72 \lesssim x_0 \lesssim 0.80$ |
| $-0.15$ | $0.72 \lesssim x_0 \leq 1.15$ | $0.72 \lesssim x_0 \lesssim 1.15$ | $\nexists x_0$ |

We set $x_0 = a_0/2M$ and $N = -(2M)^\eta/5$

**Table 4** Validity intervals of NEC, SEC and WEC for the *average cosmological constant-like fluid*. We set $x_0 = a_0/2M$ and $N = (2M)^\eta/10$

| $\lambda$ | NEC | SEC | WEC |
|---|---|---|---|
| 0 | $1.15 \lesssim x_0 \leq 1.5$ | $1.15 \lesssim x_0 \leq 1.5$ | $\nexists x_0$ |
| 0.15 | $1.15 \lesssim x_0 \leq 1.5$ | $1.15 \lesssim x_0 \leq 1.5$ | $1.15 \lesssim x_0 \lesssim 1.27$ |
| $-0.15$ | $1.15 \lesssim x_0 \leq 1.5$ | $1.15 \lesssim x_0 \leq 1.5$ | $\nexists x_0$ |

## 4 Linear stability

Wormholes are fascinating objects that deserve to be studied by their own merits [1]. As is well-known they act as tunnels connecting different parts of universe or they could connect two different universes as well. These configurations can exist in alternative gravity theories without the need of exotic matter [21–27] or even better without matter [28]. The mere existence of these theoretical solutions triggers a natural questions about their stability. In order to these wormholes truly act as a bridge between two worlds we need to guarantee their mechanical stability otherwise they can disappear and a physical observer will not have a safe passage through them. We will explore linear stability of the aforesaid wormhole solution within the context of RG by performing small perturbations which preserve the original symmetry of the configurations. In particular, Poisson and Visser [57] developed a straightforward approach for analyzing this aspect for thin-shell wormholes. The main trait of these wormholes is that their supporting matter is located on a shell placed at the joining surface; so the theoretical tools for treating them is the Darmois–Israel formalism, which leads to the Lanczos equations [2]. The solution of the Lanczos equations gives the dynamical evolution of the wormhole once an equation of state for the matter on the shell is provided. Such a procedure has been subsequently followed to examine the stability of more general spherically symmetric configurations (see, for example, Refs. [6,7,16,58–66]).

To study the stability of the system of Eqs. (40) and (41), we must consider small perturbations around their static solutions. Given (40) and (41) we confirm that the aforesaid

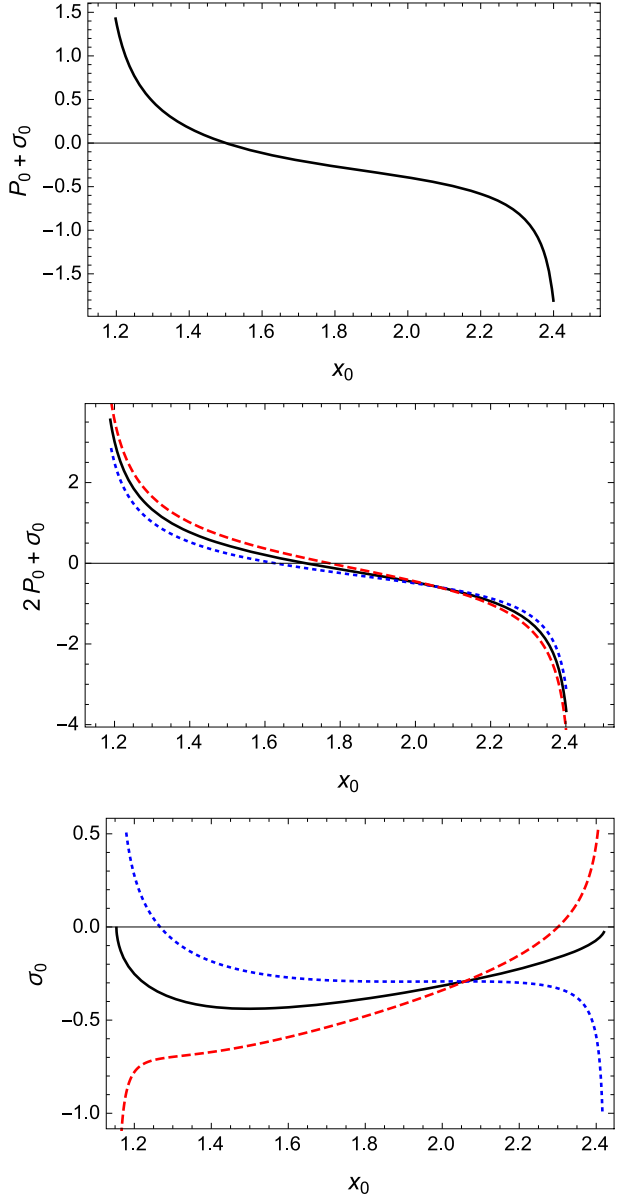

**Fig. 5** Plots of $\sigma_0 + \mathcal{P}_0$, $\sigma_0 + 2\mathcal{P}_0$ and $\sigma_0$ (in units of $M$) against $x_0 \doteq a_0/2M$ for the *average radiation-like fluid* in the extremal case, respectively. We set $N = (2M)^\eta/10$ and consider three cases: 1. $\lambda = 0$ (black, solid line), 2. $\lambda = 0.15$ (blue, dotted line), and 3. $\lambda = -0.15$ (red, dashed line)

equations satisfy the conservation equation (31). We employ the set of Eqs. (40) and (31), instead of Eqs. (40)–(41) to perform the stability analysis. Equation (40) is equivalent to the Hamiltonian constraint for the thin-shell in RG:

$$\dot{a}^2 = -\frac{16f - [a\sigma(a)]^2}{16(1-3\lambda)^2} + \frac{\lambda}{(1-3\lambda)^2} \left\{ \left[ 6f - \frac{a^2}{4} \sigma(\sigma + \mathcal{P}) \right] \right.$$
$$\left. + \lambda \left[ \frac{a^2}{4} (\sigma + \mathcal{P})^2 - 9f \right] \right\}. \tag{56}$$

Using (30), we can write a relation between $\sigma'(a)$ and $\mathcal{P}(a)$,

$$\sigma' = -\frac{2(1-3\lambda)}{1-2\lambda(1+\beta)} \frac{\sigma + \mathcal{P}}{a}, \tag{57}$$

where prime denotes derivatives with respect to $a$, that is, $\sigma' \doteq d\sigma/da$. Here we denote $\beta \doteq d\mathcal{P}/d\sigma$ being $\beta$ a free parameter. Only in the case of ordinary matter we will be able to interpret it as the square of velocity of sound, $\beta = v_{sound}^2 \in [0, 1]$. Choosing a specific equation of state, $\mathcal{P} = \mathcal{P}(\sigma)$, it is possible to solve Eq. (30) and find $\sigma(a)$. We replace that equation of state $\mathcal{P}(\sigma)$ into (57), and by doing so, $\sigma(a)$ is found. Hence, $a(\tau)$ is obtained by inserting $\sigma(a)$ in Eq. (56) and integrating by quadrature. However, we will follow another strategy to determine the wormhole's stability. The goal is to examine the dynamical evolution of Eq. (56) under small perturbations of the static solutions preserving the original symmetry. Equation (56) can be written as $\dot{a}^2 = -V(a)$ where $V(a)$ represents an effective potential. By expanding the potential $V(a)$ in a Taylor series around its static value $a_0$, we arrive at

$$\dot{a}^2 = -V(a_0) - V'(a_0)(a - a_0) - \frac{1}{2} V''(a_0)(a - a_0)^2 + \cdots \tag{58}$$

In the linear regime, we have that the static solutions, denoted by $a = a_0$, satisfy $V(a_0) = 0$ and $V'(a_0) = 0$. Discarding high order terms in the series, we may say that wormholes are stable if $V''(a_0) > 0$ (so that $V(a_0)$ is a local minimum), while for $V''(a_0) < 0$ perturbations can grow at least until the non-linear regime is reached. The situation resembles the problem of a particle moving under the influence of a given potential energy $V(a)$. If the particle is located at a local maximum, such that $V''(a_0) < 0$, then a small perturbation around its position will provoke a motion to the left or the right, such that the distance to $a_0$ will grow without limit. In order to avoid this scenario and the particle must be located at a local minimum of the potential, i.e., $V''(a_0) > 0$ (stable configuration). The second derivative of the potential can be written as

$$V''(a) = f'' - \frac{1}{8} \frac{1}{(1-3\lambda)^2 [1 - 2\lambda(\beta+1)]}$$
$$\left\{ \left[ 4\beta + 3 + 10\lambda\beta(8\lambda - 3 - 8\lambda^2) + 4\lambda(21\lambda - 20\lambda^2 - 7) \right] \sigma^2 \right.$$
$$+ 2(1-4\lambda) \left[ 2\beta + 3 + 2\lambda\beta(10\lambda - 7) + 5\lambda(4\lambda - 3) \right] \sigma\mathcal{P}$$
$$\left. + 4 \left[ 1 + 2\lambda\beta(9\lambda - 10\lambda^2 - 2) - \lambda(20\lambda^2 - 19\lambda + 7) \right] \mathcal{P}^2 \right\}. \tag{59}$$

We will evaluate (59) at the static configuration where the wormhole radius is denoted as $a_0$, thus $\dot{a}_0 = \ddot{a}_0 = 0$. Equation (59) reduces to the one found in GR when the Rastall's parameter vanishes,

$$V''[a; \lambda = 0] = f'' - \frac{1}{8} \left[ (\sigma + 2\mathcal{P})^2 + 2\sigma (1 + 2\beta) (\sigma + \mathcal{P}) \right]. \tag{60}$$

Having the second derivative of the potential at hand, we are able to determine the regions of stable configurations just by looking at which zones in the $(\beta_0, a_0)$-plane correspond to $V''(a_0) > 0$, whereas those regions associated with $V''(a_0) < 0$ are unstable. Our next task then is to analyze the stability regions given an average equations of state of the bulk fluid.

To do so, we take the solution of the Rastall field equations given by Eqs. (12) and (13) and replace in the expression for the second derivative of the potential (59). We obtain an equation that essentially depends on $\lambda, \sigma, \mathcal{P}, \beta, a, M, N$ and $\omega$. We can eliminate $\sigma$ and $\mathcal{P}$ dependence from the conditions $V(a_0) = V'(a_0) = 0$. In this way, we arrived at the final expression for $V''(a_0)$. In the following subsections, we will set the parameters $M$ and $N$ to some specific values and we will analyze the regions with $V''(a_0) > 0$ in the $(\beta_0, a_0)$-plane for different average equations of state parameter, $\omega$, describing *dust, radiation*, and *cosmological constant* fluids, respectively. Nevertheless it must be emphasized that the meaning of average equation state can be misleading and it is important to indicate the partial equation of state.

4.1 Average dust case $\omega = 0$: anisotropic fluid with $\omega_r = -1$ and $\omega_t = 1/2$

In order to be able to extract some physical information about the stability analysis, we will use same values of the Rastall parameters used in the analysis of the energy conditions. The fluid with an average dust-like equation of state corresponds to an energy-momentum $T^\mu_\nu = \rho\,\mathrm{diag}(-1, -1, 1/2, 1/2)$. As usual we consider $x = a/2M$ and $N = (2M)^\eta$ such that $f(x) \doteq 1 - x^{-1} - x^{-\eta}$. The depicted stability regions are shown in Fig. 6, where the solid area represents the region of stability. Two different values of the Rastall parameter are considered besides the value $\lambda = 0$ which represents GR.

Our aim is to compare the region of stability of RG with the ones in GR. The stability regions are severely modified due to the inclusion of $\lambda$. A positive value of the Rastall parameter enlarges the region of stability in comparison to GR while a negative value reduces the stability regions, also in comparison to GR. Along with the analysis of the energy conditions, we can see that positive values of $\lambda$ are favored. To be more clear, *it becomes possible to construct a thin-shell wormhole that has a wider stability region and at the same time it satisfies the energy conditions*. Let us mention that the $\beta_0$-line that divides two stability regions in the graphics can be found by simply computing $V''(a_0) = 0$ and it gives $\beta_0 = -1 + 1/(2\lambda)$. In the examples above, we set $\lambda = \pm 0.15$ so it leads $\beta_0 \approx 2.3$ and $\beta_0 \approx -4.3$, respectively. As it can be seen such values occur at the transition line which separates the stable region from the unstable one (cf. Fig. 6).

4.2 An average radiation-like fluid with $\omega = 1/3$: anisotropic fluid with $\omega_r = -1$ and $\omega_t = 1$

As we have done before we will split the analysis into cases: 1.—extremal and 2.—non-extremal. Notice that he metric tensor does not depend on the parameter $\lambda$ since $\eta(\omega = 1/3) = 2$.

*4.2.1 Extremal and non-extremal cases*

For the extremal case we set $x = a/2M$ and $N = -\frac{1}{4}(2M)^\eta$ such that $f(x) = 1 - x^{-1} + \frac{1}{4}x^{-2}$. Although the metric coefficient is similar to the extremal Reissner–Nordström case the junction conditions that we are using are different, therefore we could expect some modifications in the stability regions for $\lambda \neq 0$. Figure 7 displays the stability region for $\lambda = 0, \pm 0.15$. In fact, the inclusion of the Rastall parameter changes radically the stability zones provided they are enlarged for positive values of the Rastall parameter and they are reduced for negative values of $\lambda$. Then, we obtained a similar scenario that in the average dust-like case. Interestingly, it is possible to have $\beta_0$ considerably small for wormholes with small radii, namely, the wormhole could be supported by normal nonrelativistic matter.

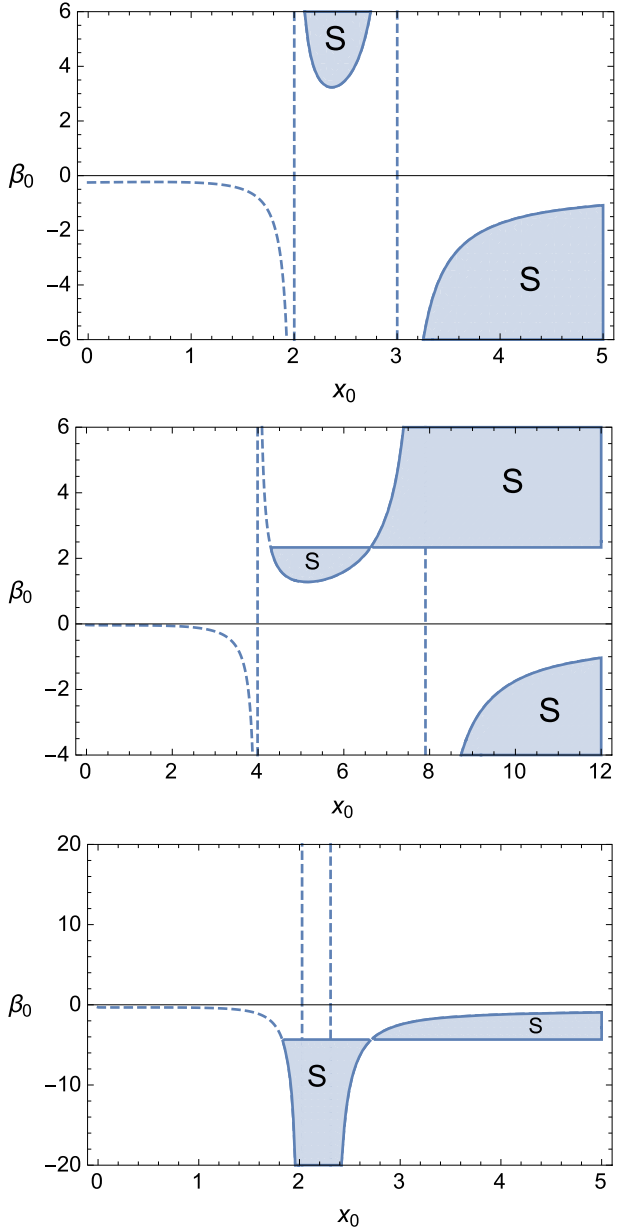

**Fig. 6** Curve $V''(a_0) = 0$ and stability regions ($V''(a_0) > 0$) for $\lambda = 0$, $\lambda = 0.15$, and $\lambda = -0.15$ in the case of an average dust-like fluid ($\omega = 0$). The horizon for $\lambda = 0$ is located at $x_h = 2$, for $\lambda = 0.15$ at $x_h \approx 4.29$ and for $\lambda = -0.15$ at $x_h \approx 1.83$

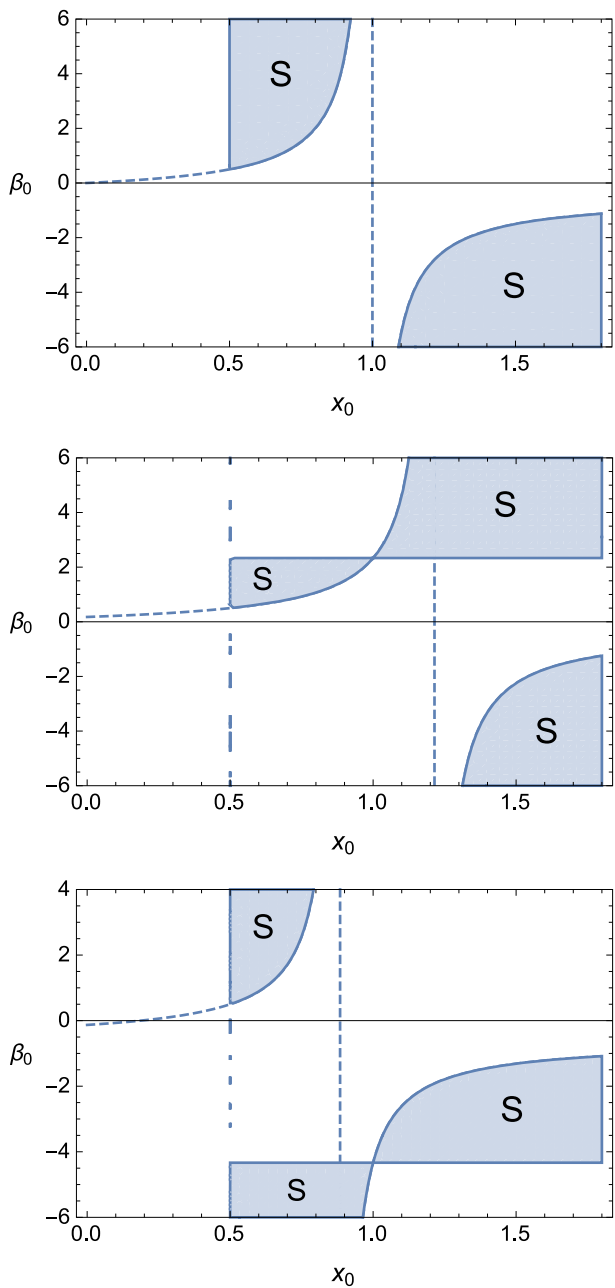

**Fig. 7** Curve $V''(a_0) = 0$ and stability regions ($V''(a_0) > 0$) for $\lambda = 0$, $\lambda = 0.15$ and $\lambda = -0.15$ for a fluid with an average equation of state $\omega = 1/3$ within the extremal case. The horizon is located at $x_h = 0.5$

For the non-extremal case, we fix $x = a/2M$ and $N = -\frac{1}{5}(2M)^\eta$ in such a way that the metric coefficient reads $f(x) = 1 - x^{-1} + \frac{1}{5}x^{-2}$. Hence, the original manifold has two horizons located at $x_h^- \approx 0.27$ and $x_h^+ \approx 0.72$, respectively. The wormhole radius must be larger than 0.72. The inclusion of the Rastall parameter modifies the stability regions by favoring positive values of $\lambda$. As it happens in the extremal cases, the RG allows us to have wormholes supported by normal nonrelativistic matter where the relation $\beta_0 = v_{sound}^2 \ll 1$ holds.

We mention in passing the results of an average cosmological constant-like fluid case with $\omega = -1$. Fixing $x = a/2M$ and $N = \frac{1}{10}(2M)^\eta$, the metric coefficient reads $f(x) = 1 - x^{-1} - \frac{1}{10}x^2$. The parameter $\eta$ does not depend on $\lambda$. The event horizon is placed at $x_h \approx 1.15$ and the cosmological horizon is at $x_h \approx 2.42$. Due to the existence of cosmological horizon in the original manifold we ended up with wormholes with finite radii, that is, $x_0 \in (1.15, 2.42)$. To conclude this section, let us mention that the above findings are a generalization of the results reported by Lobo and Crawford [67] for the Schwarzschild–de Sitter case. It becomes evident the fundamental role played by the conservation laws in the stability analysis. Despite the fact that the metric has the same functional dependence with the radius the nonconservation of the energy-momentum, which is at the core of the Rastall approach, leads to a completely different physical situations provided the stability regions are now enlarged for $\lambda > 0$.

## 5 Summary

We studied the effect of the Rastall theory of gravity [29] in the construction of thin-shell wormholes using the "cut and paste" technique proposed by Visser and Poisson some years ago [2,3,57]. In doing so, we have obtained the generalized Darmois-Israel junction conditions [47,51] within the framework of RG. One particular trait is that the new junction conditions led to a modified conservation equation for the energy-momentum associated with the matter located at the wormhole throat.

We analyzed possible modifications of the null, strong and weak energy conditions due to the Rastall parameter. We found important effects regarding the sign taken by $\lambda$. For all bulk fluids surrounded the wormhole throat, we found that a positive Rastall parameter increases the regions in which the energy conditions are satisfied, allowing in this to preserve the weak energy condition for a positive $\lambda$.

Besides, we were able to analyze the stability conditions under linear perturbations which preserve the original spherical symmetry of the spacetime [57]. We have examined the impact of the Rastall parameter when the average equation of state takes different values.

Regarding the stability analysis, we found that the stability regions are considerably modified due to the non-trivial dependence of the potential with the Rastall parameter. As a matter of fact, we showed that the Rastall parameter had a considerable impact on the regions of stability even though we took small values of $\lambda$ in all the explored cases. Positive values of $\lambda$ increased the stability regions whereas negative values decreased them in comparison to General Relativity. It should be remarked that the values of the Rastall parameter considered here are consistent with the ones obtained from a statistical analysis of 118 galaxy–galaxy strong gravitational lensing systems within the context of RG [41]. Remarkably, we found that for the radiation case (extremal and non-extremal cases) is possible to have thin-shell wormholes supported by nonrelativistic matter provided the relation $\beta_0 = v_{sound}^2 \ll 1$ holds.

We can summarize our main findings as follows. Positive values of the Rastall parameter helped to preserve the energy conditions, allowing for a non-exotic matter content at the wormhole throat, and it also enlarged the stability regions in the case of an average dust-like, radiation-like, and a cosmological constant equation of state. However, negative values of $\lambda$ reduced the allowed wormhole radii where the null and strong energy conditions are held (and kept the weak energy condition violated). In addition, it also diminished the wormhole stability regions.

**Acknowledgements** M.G.R is supported by FAPES/CAPES grant under the PPGCosmo Fellowship Programme. I. P. Lobo and J. P. M. G. are supported by CAPES. This work was partially supported by the Coordenação de Aperfeiçoamento de Pessoal de Nível Superior—(CAPES)—Finance Code 001. The work of H. Moradpour has been financially supported by Research Institute for Astronomy and Astrophysics of Maragha (RIAAM).